\documentclass[conference]{IEEEtran}
\IEEEoverridecommandlockouts

\usepackage[utf8]{inputenc}

\usepackage{amsmath,amssymb,amsfonts}
\usepackage{graphicx}
\usepackage{booktabs}
\usepackage{multirow}
\usepackage{float}
\usepackage{algorithmic}
\usepackage{adjustbox}
\usepackage{lscape}
\usepackage{textcomp}
\usepackage{xcolor}
\usepackage{tikz}
\usetikzlibrary{arrows.meta,positioning}

\usepackage{cite}
\usepackage{listings}
\usepackage{fvextra}

\usepackage[most]{tcolorbox}

\usepackage{hyperref}

\lstdefinestyle{jsoncolorkeys}{
  basicstyle=\ttfamily\small,
  breaklines=true,
  showstringspaces=false,
  literate= *{:}{{\textcolor{black}{:}}}1
           {"system"}{{\textcolor{blue}{"system"}}}1
           {"user"}{{\textcolor{purple}{"user"}}}1
           {"assistant"}{{\textcolor{teal}{"assistant"}}}1
}

\def\BibTeX{{\rm B\kern-.05em{\sc i\kern-.025em b}\kern-.08em
    T\kern-.1667em\lower.7ex\hbox{E}\kern-.125emX}}

\begin{document}

\title{Automated Bug Triaging using Instruction-Tuned Large Language Models\\}

    \author{
    Kiana Kiashemshaki\textsuperscript{1}, 
    Arsham Khosravani\textsuperscript{2},
    Alireza Hosseinpour\textsuperscript{1},  Arshia Akhavan\textsuperscript{1} \\
    \textsuperscript{1}Department of Computer Science, Bowling Green State University, Bowling Green, OH, USA \\
       Emails: \{kkiana, Arshiaa, Ahosein\}@bgsu.edu \\
       \textsuperscript{2}Department of Electrical and Computer Engineering,
       California State University Northridge, CA, USA \\
       Email: arsham.khosravani.085@my.csun.edu
}

\maketitle

\begin{abstract}
Manual bug triaging is slow, inconsistent, and increasingly unsustainable in large-scale software projects. We present an instruction-tuned, project-specific large language model (LLM) for automated bug triaging that integrates \textbf{candidate-constrained decoding} to guarantee valid developer assignments. Our framework fine-tunes a LoRA adapter on the 8B-parameter \emph{DeepSeek-R1-Distill-Llama-8B} model using a conversational JSONL format derived from EclipseJDT and Mozilla issue-tracker data. The approach requires no handcrafted features, graph construction, or complex preprocessing, enabling rapid adaptation to new projects with minimal engineering effort. At inference time, candidate-constrained decoding restricts outputs to a known roster and produces ranked Top-$K$ developer recommendations. Using the \textbf{same multi-year temporal window and filtering protocol} as prior graph-based work, our model attains \textbf{Top-1/Hit@10 of 0.156/0.475 on EclipseJDT} and \textbf{0.013/0.753 on Mozilla}. The method is compute-efficient leveraging 4-bit NF4 quantization and fully reproducible, with code, prompts, and per-issue predictions available for replication. These results show that lightweight, instruction-tuned LLMs can generate practical, high-quality shortlists for real-world bug triaging and motivate hybrid extensions to improve exact Top-1 assignment.

\footnote{All code, prompts, and predictions will be released in an open repository.}

\end{abstract}


\section{Introduction}

In software development, when a new issue is reported through a project’s issue tracking system (e.g., Jira, Bugzilla, or GitHub), it is typically reviewed by a project maintainer also known as a triager who verifies the report, identifies duplicates, sets its severity, and assigns it to the appropriate developer~\cite{wu2022spatial, zhou2025issuecourier}. This process, known as \emph{bug triaging}, plays a critical role in ensuring timely and accurate issue resolution. When triaging is ineffective, it can lead to \emph{bug tossing}, where issues are repeatedly reassigned before reaching the EclipseJDT project, some bugs have required more than a dozen reassignments and taken up to 100 days to resolve~\cite{wu2022spatial}, significantly increasing maintenance costs and delaying fixes~\cite{yadav2024developer, samir2023interpretable}.

Effective triaging is inherently challenging. It requires deep knowledge of the project, its codebase, and the expertise and availability of contributors~\cite{hajari2024sofiawl, aini2025expertise}. These challenges are especially pronounced in large-scale open-source projects, where contributor roles are fluid and expertise is not always clearly defined, making manual bug assignment error-prone~\cite{zhou2025issuecourier}.

As software projects have scaled, the number of reported issues has grown dramatically, making manual bug triaging increasingly unsustainable. Projects like EclipseJDT, Mozilla, and Google Chromium receive thousands of new reports each month~\cite{xie2021devrec, zhou2025issuecourier}. This volume has driven interest in automated bug triaging solutions, which can be broadly grouped into two categories~\cite{dong2024neighborhood}:  
(1) text classification-based approaches and (2) graph-based approaches.

Text classification-based methods treat bug triaging as a supervised classification problem, using textual elements of bug reports (e.g., titles, descriptions) as features and the assigned developer as the label. Recent transformer-based models have shown strong results by capturing contextual semantics~\cite{arnob2025bug, dipongkor2023comparative, lee2022light, wang2024empirical}. However, these models face key limitations:  
(i) incorporating all available bug text can introduce noise~\cite{zhang2023ealink}, preventing convergence, and  
(ii) transformer models have token span limits that restrict how much context they can use~\cite{akhavan2025gitanchor}.  
Moreover, report structures vary across projects, impacting generalizability~\cite{zhou2025issuecourier, zhang2023ealink, arnob2025bug}.

Despite these advances, relying solely on textual features can be insufficient. To address this, researchers have explored graph-based methods that model relationships between bugs, developers, and other artifacts~\cite{zhou2025issuecourier, cao2025complex, wu2022spatial, tao2025structural, dong2024neighborhood, dai2023graph}. Graph neural networks (GNNs)~\cite{wu2020comprehensive} learn node embeddings that capture structural, temporal, and semantic correlations, leveraging the insight that a developer’s expertise is reflected in the bugs they have resolved~\cite{wu2022spatial}. While effective, these methods can be computationally expensive, rely on partial graph sampling (e.g., random walks~\cite{wu2022spatial}), and depend on explicit bug–developer links, which are often sparse or noisy~\cite{dong2024neighborhood}.

A common limitation across both categories is their reliance on costly retraining and feature engineering to keep up with evolving developer activity and collaboration patterns~\cite{dong2024neighborhood, wu2022spatial, dipongkor2023comparative, lee2022light, arnob2025bug}. They also assume that the listed assignee is the actual resolver, despite evidence of frequent discrepancies, leading to noisy labels~\cite{zhou2025issuecourier}.

While prior work has shown strong results with transformer encoders and graph-based models, LLMs for triaging are either used zero-shot or as components inside larger pipelines, with limited evidence on instruction-tuned, project-specific models that \emph{constrain} outputs to valid assignees. The impact of such candidate-constrained decoding on ranked recommendations (Hit@K) has not been systematically studied under large, multi-year project datasets.

In this work, we investigate the use of large language models (LLMs) for bug triaging. We construct datasets from the EclipseJDT and Mozilla projects, following the same multi-year temporal windows and filtering protocol introduced in NCGBT~\cite{dong2024neighborhood}. After cleaning and formatting them into JSONL conversational prompts, we fine-tune a LoRA adapter on top of the DeepSeek-R1-Distill-Llama-8B model. During inference, we apply candidate-constrained decoding to restrict outputs to valid assignee identifiers, producing a ranked Top-$K$ list.

We report Top-1 and Hit@K results for both projects and position these findings in direct comparison with prior text classification- and graph-based methods. Our results highlight the scalability and low engineering overhead of LLM-based triaging in practice.

\begin{figure}[htbp]
\centering
\includegraphics[width=0.35\textwidth]{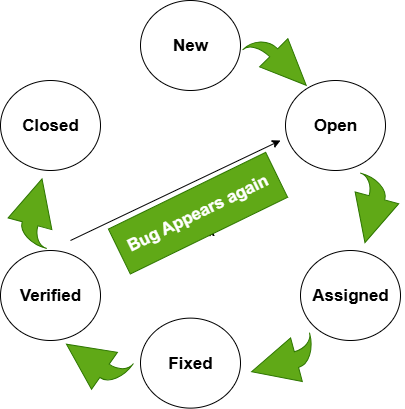}
\caption{States of a bug. Accurate triaging prevents repeated reassignment and delays.}
\label{fig:bug_states}
\end{figure}

The remainder of this paper is organized as follows: Section~\ref{sec:related} reviews related work, Section~\ref{sec:method} details the methodology, Section~\ref{sec:eval} presents the results, and Section~\ref{sec:concl} concludes the paper. Figure~\ref{fig:bug_states} illustrates the typical lifecycle of a bug.

\section{Related Work}
\label{sec:related}

Automatic bug triaging plays a pivotal role in modern software maintenance workflows by reducing manual effort and accelerating the resolution of reported defects~\cite{dipongkor2023comparative, lee2022light}. Over the years, researchers have proposed a range of techniques to automate the bug assignment process. In this section, we categorize prior work into \emph{text classification-based}, \emph{graph-based}, and \emph{LLM-based} methods, and discuss each in detail.

\subsection*{Text Classification-Based Methods}
These approaches typically treat elements of bug reports (e.g., titles and descriptions) as structured text input. This text is transformed into feature representations and used to train machine learning (ML) or deep learning (DL) models, with developers treated as classification labels. Earlier work relied on classic ML techniques such as Naive Bayes~\cite{murphy2004automatic}, Support Vector Machines (SVM)~\cite{anvik2006automating}, Term Frequency Inverse Document Frequency (TF–IDF)~\cite{xuan2014towards}, and Latent Dirichlet Allocation (LDA)~\cite{xia2016improving}. However, these methods have limited semantic understanding of textual features and struggle to capture the contextual nuances of bug report information~\cite{cao2025complex, zhou2025issuecourier, zhang2023ealink, arnob2025bug}.

Recent studies address this limitation by leveraging deep learning models capable of capturing context-dependent semantic features~\cite{arnob2025bug, dipongkor2023comparative, lee2022light}. Lee et al.~\cite{lee2022light} proposed a lightweight framework that fine-tunes pre-trained models like RoBERTa~\cite{liu2019roberta} and DeBERTa~\cite{he2020deberta} on bug reports, achieving competitive accuracy with lower computational cost. Dipongkor et al.~\cite{dipongkor2023comparative} compared transformer-based models for bug triaging across several open-source datasets, showing that DeBERTa~\cite{he2020deberta} outperformed BERT~\cite{devlin2019bert} and RoBERTa~\cite{liu2019roberta}. Wang et al.~\cite{wang2024empirical} evaluated 35 deep learning-based bug assignment models, combining five embeddings and seven classifiers, and found Bi-LSTM with attention or ELMo embeddings to be consistently strong. Mani et al.~\cite{mani2019deeptriage} combined a bidirectional recurrent neural network with an attention mechanism, using Word2Vec~\cite{mikolov2013efficient} for representation learning. Aung et al.~\cite{aung2022multi} introduced a multi-task model that jointly assigns developers and classifies issue types, combining a CNN-based text encoder with a BiLSTM-based AST encoder. Arnob et al.~\cite{arnob2025bug} used XLNet for bug triaging and showed that incorporating commit messages with summaries and descriptions significantly improves accuracy.

Despite their effectiveness, text classification-based approaches remain limited by:  
(i) the risk of introducing noise and bloating the dataset when incorporating all available text~\cite{zhang2023ealink}, and  
(ii) architectural constraints in transformer models, such as short context windows and limited token spans~\cite{akhavan2025gitanchor}.  
To mitigate these, many methods rely on selective sampling, term selection, or manual feature engineering. However, the quality and length of available attributes vary significantly across projects~\cite{zhou2025issuecourier, zhang2023ealink, arnob2025bug}, weakening generalization.  

In summary, while text classification methods have advanced significantly with transformers, their reliance on limited fields, sensitivity to noise, and cross-project generalization challenges motivate the need for more flexible solutions.

\subsection*{Graph-Based Methods}
To overcome the limitations of relying solely on textual information, recent studies have turned to graph-based methods that model the relationships between bugs, developers, and other relevant entities~\cite{zhou2025issuecourier, wu2022spatial, cao2025complex, dai2023graph, dong2024neighborhood}. These approaches are grounded in the intuition that a developer’s expertise is reflected in their past issue-resolution activity. Graph Neural Network (GNN) architectures~\cite{wu2020comprehensive} are applied to learn node embeddings that capture both structural and semantic content.  

GCBT~\cite{dai2023graph} constructs a bipartite bug–developer graph with bug nodes initialized using a GRU-based NLP module, and embeddings learned through GNNs. NCGBT~\cite{dong2024neighborhood} enriches node representations using contrastive learning over structural and semantic neighbors. SCL-BT~\cite{tao2025structural} employs self-supervised edge perturbation and hypergraph sampling to capture both labeled and unlabeled bug–developer associations.

Recent methods also incorporate temporal dynamics~\cite{zhou2025issuecourier, wu2022spatial, cao2025complex}, recognizing that developer activity changes over time~\cite{jahanshahi2023adptriage}. ST-DGNN~\cite{wu2022spatial} models periodic developer behavior at multiple time scales using graph recurrent convolutions. IssueCourier~\cite{zhou2025issuecourier} formalizes five relationship types and slices graphs over time to model evolving developer behavior. Cao et al.~\cite{cao2025complex} focus on inter-bug dependencies, modeling blocking relationships in time-ordered snapshots.

Early graph-based methods suffer from high computational costs and scalability limits, leading many to adopt random walk sampling~\cite{wu2022spatial, grover2016node2vec, nguyen2018continuous}, which can miss important patterns. They also depend on explicit bug–developer links, often sparse or noisy~\cite{tao2025structural, dong2024neighborhood}.  

In summary, graph-based methods capture richer structural and temporal relationships than text alone, but require complex graph construction, are sensitive to noisy interactions, and incur high computational overhead.

\subsection*{LLM-Based Methods}
The emergence of Large Language Models (LLMs) such as GPT-4, Claude, and LLaMA~\cite{touvron2023llama, openai2024gpt4} has opened new possibilities for automated bug triaging. Unlike static transformer-based encoders~\cite{lee2022light, dipongkor2023comparative, arnob2025bug}, LLMs are trained on massive, diverse corpora and instruction-tuned to follow natural language prompts. This allows them to process heterogeneous bug content including long descriptions, code snippets, and discussion threads without requiring handcrafted feature extraction or strict field selection.

Only a few recent studies have explored LLMs in this task. Amini et al.~\cite{amini2024llmbt} showed that zero-shot prompting with GPT-4 can achieve competitive Top-$k$ accuracy versus fine-tuned BERT-based baselines, though results are sensitive to prompt design and dataset noise. Zhou et al.~\cite{zhou2025issuecourier} explored generative models for developer recommendation but as components in larger graph-based pipelines, rather than as standalone triaging systems.  

These early works suggest LLMs can adapt to evolving project contexts without retraining, but systematic evaluation against state-of-the-art classification and graph-based approaches is lacking.Our work extends this direction with an instruction-following LLM fine-tuned on cleaned and structured EclipseJDT and Mozilla datasets, using the same multi-year temporal windows and filtering protocol as NCGBT~\cite{dong2024neighborhood}. This approach directly maps natural language bug reports to developer assignments, leveraging the LLM’s capacity for long-context reasoning and multi-format input understanding, and enables a direct comparison with prior transformer- and graph-based methods.

\section{Methodology}
\label{sec:method}

This section documents the end-to-end pipeline we implemented for automated issue triage using a LoRA-adapted large language model (LLM). We first describe the construction of two project-specific datasets from the EclipseJDT and Mozilla issue trackers following the NCGBT multi-year temporal windows. We then formalize the task (single-label assignee prediction with Top-$K$ retrieval), detail the base model and adapter configuration (DeepSeek-R1-Distill-Llama-8B with LoRA), and specify the supervised fine-tuning (SFT) procedure, decoding strategy, and candidate-constrained ranking used to compute Hit@$K$. Finally, we present the evaluation metrics (Top-1 accuracy, Hit@$K$), implementation details, and baselines required for exact reproducibility.

\subsection{Dataset Collection}
Following the setup of Dong et al.~\cite{dong2024neighborhood}, we curated two large-scale, multi-year datasets: EclipseJDT and Mozilla.  

EclipseJDT: This dataset originates from the EclipseJDT project and contains detailed information on approximately 20{,}000 resolved bugs and their associated developers. Each record includes the bug description, summary, fixer, priority, status, and other metadata. The time span covers November 2007 to November 2015.  

Mozilla: This dataset is derived from the Mozilla project and contains around 120{,}000 resolved bug reports, covering the period from June 1999 to February 2021. Each record provides the same fields as EclipseJDT.  

To ensure data quality and consistency with NCGBT, we removed developers with fewer than 10 resolved bugs along with their corresponding reports. After cleaning, we randomly split each dataset into 80\% training, 10\% validation, and 10\% test sets.  

\begin{table}[t]
\centering
\caption{Dataset Details}
\label{tab:dataset_details}
\setlength{\tabcolsep}{8pt}
\renewcommand{\arraystretch}{1.2}
\begin{tabular}{|l|r|r|r|r|}
\hline
\textbf{Datasets} & \textbf{Bugs} & \textbf{Developers} & \textbf{Relationships} & \textbf{Density} \\
\hline
EclipseJDT & 16106 & 4017 & 53985 & 0.0008 \\
Mozilla    & 110467 & 37371 & 569289 & 0.0001 \\
\hline
\end{tabular}
\end{table}

Each issue was reformatted into line-delimited JSON (JSONL) for compatibility with instruction-tuned LLMs. Each JSONL record contains three roles:
\begin{itemize}
  \item System: framing the model as an expert bug triager.
  \item User: concatenation of the issue title and full description.
  \item Assistant: the assigned developer’s email or name.
\end{itemize}

\begin{figure}[h]
  \centering
  \includegraphics[width=0.95\linewidth]{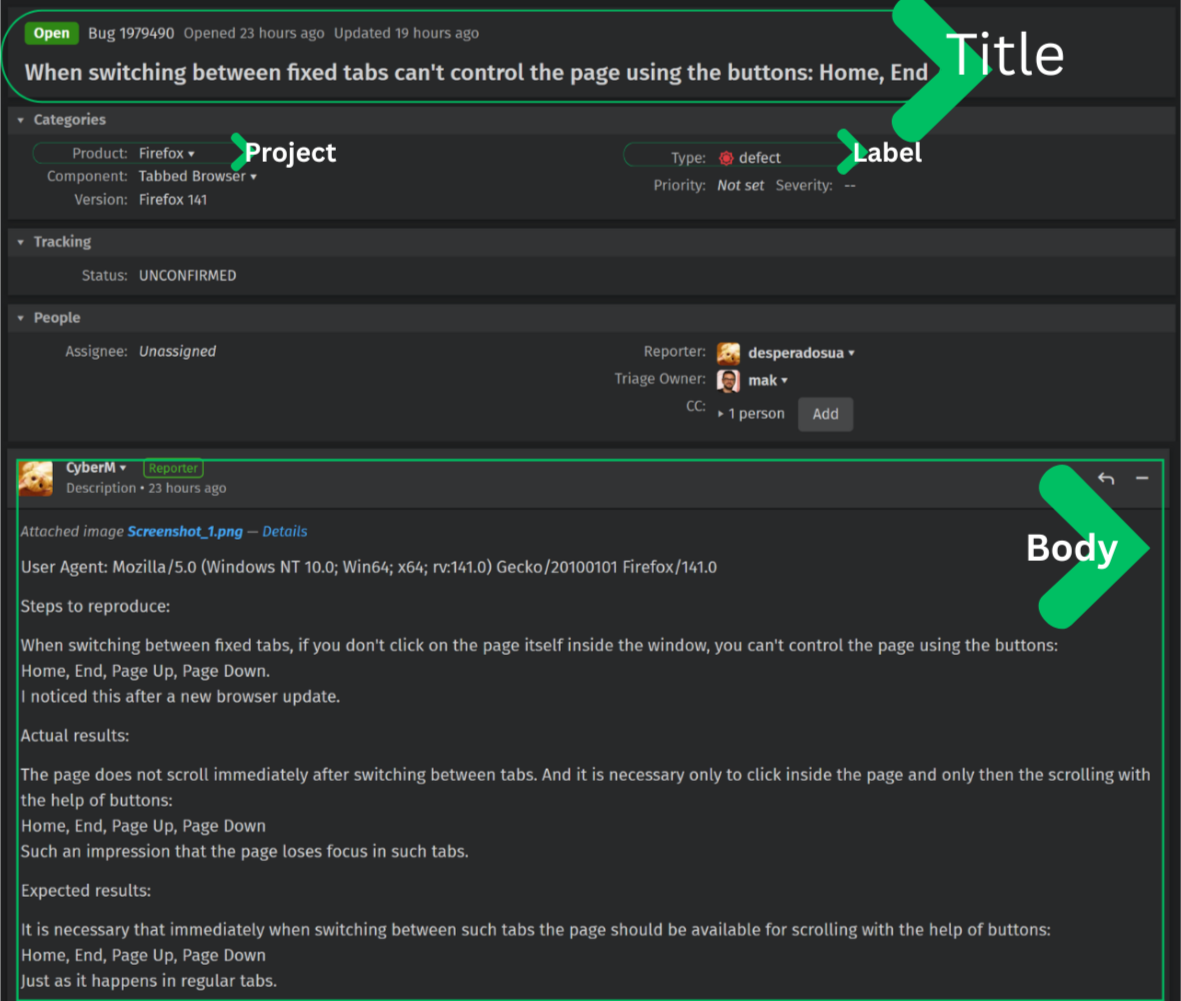}
  \caption{Example of a Mozilla bug report, with key elements used in prompt creation.}
  \label{fig:issue-structure}
\end{figure}

This conversational format matches the input style of modern LLMs, enabling natural prompt response learning and direct comparison with NCGBT results.

\begin{table}[h]
\centering
\scriptsize
\caption{Dataset statistics for EclipseJDT and Mozilla bug reports}
\label{tab:dataset_stats}
\begin{tabular}{lcccc}
\toprule
\textbf{Project} & \textbf{Total Issues} & \textbf{Train (80\%)} & \textbf{Validation (10\%)} & \textbf{Test (10\%)} \\
\midrule
EclipseJDT & $\sim$20{,}000 & 16{,}000 & 2{,}000 & 2{,}000 \\
Mozilla    & $\sim$120{,}000 & 96{,}000 & 12{,}000 & 12{,}000 \\
\bottomrule
\end{tabular}
\end{table}

\begin{tcolorbox}[
  colback=gray!10,
  colframe=gray!80,
  title=Example JSONL Record for LLM Fine-Tuning,
  fontupper=\ttfamily\small,
  sharp corners,
  boxrule=0.5pt,
  arc=4pt,
  breakable,
  enhanced
]
{
  "system": "You are an expert bug triager.",\\
  "user": "Title: App crashes when saving.\\nDescription: Steps to reproduce...",\\
  "assistant": "dev@example.com"
}
\end{tcolorbox}

The datasets preserve real-world vocabulary, formatting styles, and developer assignment patterns, ensuring realistic evaluation.

\subsection{Problem Formulation}

We formulate automated issue triaging as a single-label classification problem.  
Given a Bugzilla issue $x$ consisting of a \emph{title} and \emph{body} text, the goal is to predict the correct assignee $y$:

\[
f_{\theta} : \{\text{title}, \text{body}\} \rightarrow \text{assignee}
\]
where $f_{\theta}$ is a supervised model parameterized by $\theta$.

In addition to the primary Top-1 Accuracy metric, we also evaluate Top-$K$ retrieval accuracy (\text{Hit@K}), which measures whether the true assignee appears within the first $K$ ranked suggestions generated by the model.

\subsection{Prompt Design and Data Shaping}

\subsubsection{Training Prompt (SFT)}
Each training example was embedded into a strict instruction template to ensure consistent task framing and minimize output drift:

\lstset{
    basicstyle=\ttfamily,
    breaklines=true,
    columns=fullflexible,
    frame=single,
    xleftmargin=0pt,
    xrightmargin=0pt
}
\begin{lstlisting}
Below is an issue. Suggest the single best developer to resolve it. 

### Issue:
{title + "\n\n" + body}

### Assignee: {gold_assignee}
\end{lstlisting}

During SFT, the LoRA-adapted model was trained to produce only the assignee identifier following the \texttt{\#\#\# Assignee:} anchor.

\subsubsection{Inference Prompt (Top-1)}
The same template was used at inference time, but without the gold label, ensuring consistency between training and prediction.

\subsubsection{Inference Prompt (Top-K)}
To compute Hit@$K$ and reduce hallucinations of non-existent identifiers:
\begin{itemize}
    \item A candidate list of valid assignees was provided, derived from training labels or the project's official roster.
    \item The model was instructed to output:
\begin{verbatim}
Top 10 unique assignees, comma-separated,
no extra words.
\end{verbatim}
    \item Outputs were post-processed using:
    \begin{enumerate}
        \item A strict regular expression to validate email or handle format.
        \item Deduplication while preserving the original order.
        \item Padding with \texttt{None} if fewer than $K$ valid suggestions remained.
    \end{enumerate}
\end{itemize}

All metrics reported in this paper use candidate lists built \emph{only} from training labels and/or an official roster, avoiding test-label leakage.

\subsection{Model Architecture and LoRA Configuration}

The base model was \texttt{DeepSeek-R1-Distill-Llama-8B}, loaded with 4-bit NF4 quantization for parameter storage while performing computation in \texttt{float16} when hardware support was available. This significantly reduced GPU memory usage without degrading numerical stability during training.

A LoRA (Low-Rank Adaptation) adapter was applied to the following transformer projection modules:
\begin{itemize}
    \item \texttt{q\_proj}, \texttt{k\_proj}, \texttt{v\_proj}, \texttt{o\_proj}
    \item \texttt{gate\_proj}, \texttt{up\_proj}, \texttt{down\_proj}
\end{itemize}

LoRA hyperparameters were:
\begin{itemize}
    \item Rank ($r$): 16
    \item Scaling factor ($\alpha$): 16
    \item Dropout: 0.0
\end{itemize}

By freezing the base model weights and learning only low-rank updates, LoRA enabled efficient fine-tuning with minimal computational overhead, reducing both training cost and memory footprint while retaining the original model’s representational capacity.

\subsection{Training Procedure}

We trained the model via SFT with a cross-entropy loss on the next-token prediction task, using the following configuration:
\begin{itemize}
    \item Batch size: 2 (gradient accumulation over 4 steps for an effective batch size of 8).
    \item Maximum steps: 500 (approximately one full epoch over the training set).
    \item Learning rate: $2 \times 10^{-4}$; weight decay: 0.01.
    \item Warmup ratio: 0.03 of total steps.
    \item Optimizer: AdamW.
    \item Scheduler: Linear learning rate decay.
    \item Sequence length: 2048 tokens.
    \item Random seed: 3407 (for reproducibility).
    \item Checkpoints: both intermediate and final LoRA adapter weights saved.
\end{itemize}

This setup enabled efficient fine-tuning within limited computational resources while ensuring stable convergence and minimizing overfitting.

\section{Evaluation}
\label{sec:eval}

This section describes our evaluation setup on the NCGBT-aligned multi-year datasets
(EclipseJDT: Nov 2007–Nov 2015; Mozilla: Jun 1999–Feb 2021). Unless otherwise
stated, all metrics are computed on the cleaned test splits after developer filtering
(<10 removals), with the label space restricted to developers observed in training
(no test-label leakage). For EclipseJDT the test set has 2{,}000 issues; for Mozilla
the test set has 12{,}000 issues. We report exact-match Top-1 accuracy and Hit@$K$
($K\!\in\!\{1,\dots,10\}$) from candidate-constrained decoding.

\subsection{Evaluation Protocol}

\paragraph{Inputs and outputs.}
Each test instance consists of a single issue composed of \emph{title + body}. The model generates (i) a single identifier for Top-1, or (ii) a ranked list of up to ten identifiers for Hit@K, constrained to a candidate set built from training labels.

\paragraph{Top-1 Accuracy.}
For each issue, the title and body were concatenated as:
\begin{verbatim}
title + "\n\n" + body
\end{verbatim}
Predictions were extracted from the model output after the {\#\#\# Assignee:}  anchor. Top-1 accuracy is the fraction of exact matches between the predicted assignee and the gold-standard assignee.

\paragraph{Hit@K}
Following prior work~\cite{dong2024neighborhood}, Hit@K is the fraction of test issues for which the true assignee appears within the top-$K$ predictions:
\[
  \mathrm{Hit@}K = \frac{N_{\text{hit}}}{N_{\text{pred}}}.
\]

\subsection{Example Predictions}

\textit{Case 1 – Correct Assignment}  
System: \texttt{You are an expert bug triager.}  
User: \texttt{Title: App crashes when saving. Description: Clicking save on an empty file causes a crash.}  
Assistant (Ground Truth): \texttt{alice.dev@mozilla.org}  
Assistant (Model Output): \texttt{alice.dev@mozilla.org}  

\textit{Case 2 – Cold-Start Failure}  
System: \texttt{You are an expert bug triager.}  
User: \texttt{Title: UI glitch when resizing window. Description: Rapid resizing causes overlapping UI icons.}  
Assistant (Ground Truth): \texttt{dev2@mozilla.org}  
Assistant (Model Output): \texttt{dev1@mozilla.org}  

These examples illustrate both accurate predictions and cold-start limitations when the target developer had little or no representation in training.

\subsection{Workflow}
\begin{figure}[h]
    \centering
    \includegraphics[width=0.85\linewidth]{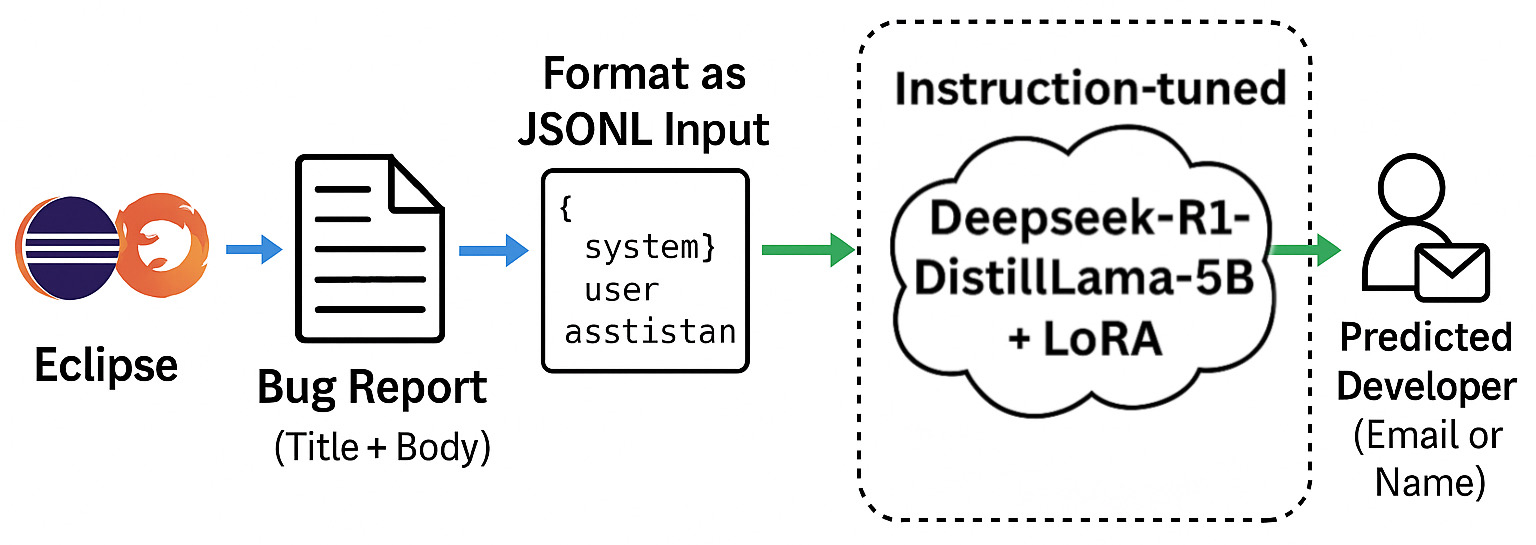}
    \caption{Our LLM-based bug triaging workflow.}
    \label{fig:workflow}
\end{figure}

Figure~\ref{fig:workflow} summarizes the complete pipeline from data collection and JSONL formatting to model training and prediction.

\subsection{Main Results}

\begin{figure*}[t]
  \centering
  \includegraphics[width=\textwidth]{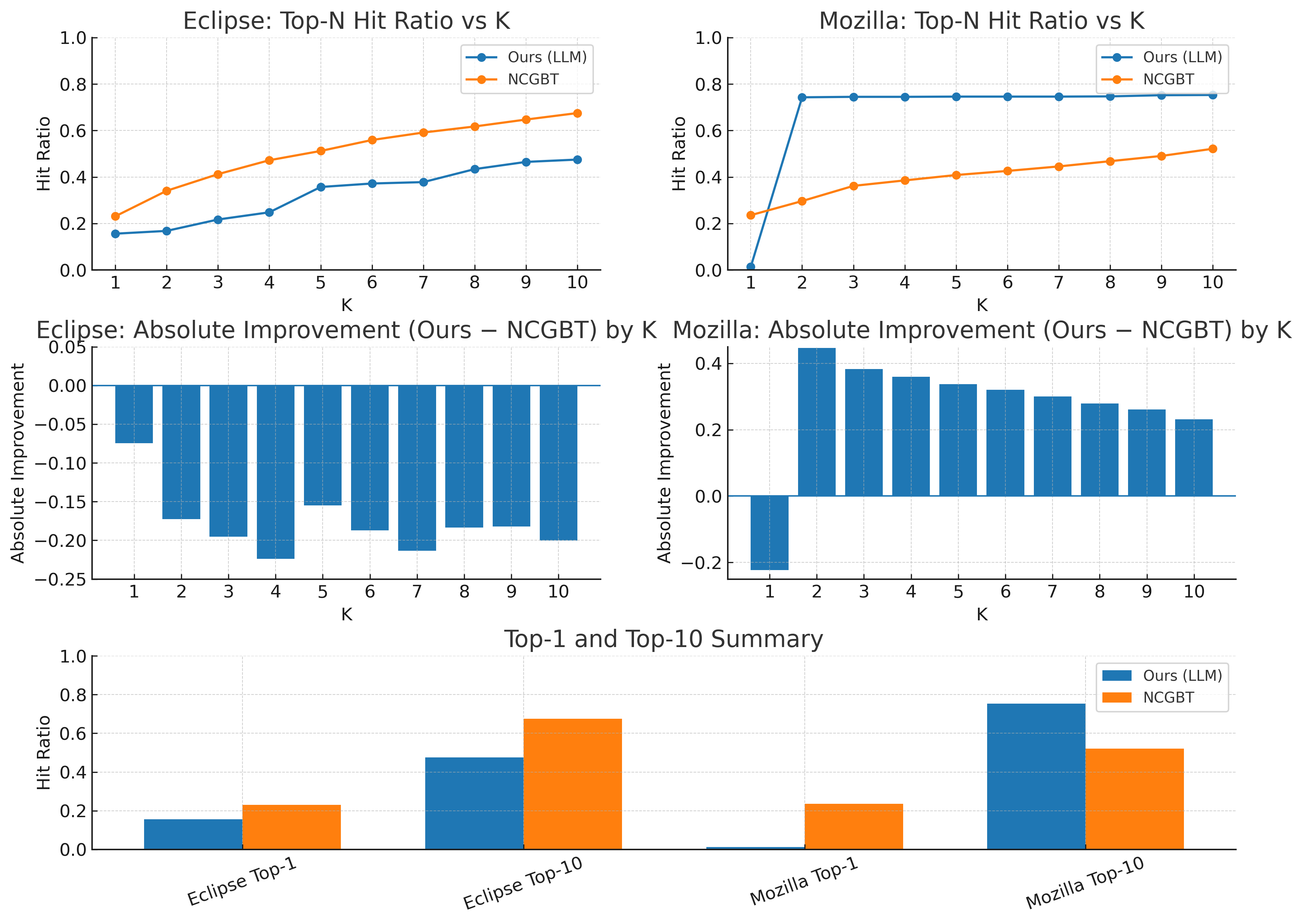}
  \caption{EclipseJDT and Mozilla: Hit@K curves for our LLM vs.\ NCGBT (multi-year evaluation).}
  \label{fig:combined_results}
\end{figure*}

\begin{figure}[htbp]
    \centering
    \includegraphics[width=\linewidth]{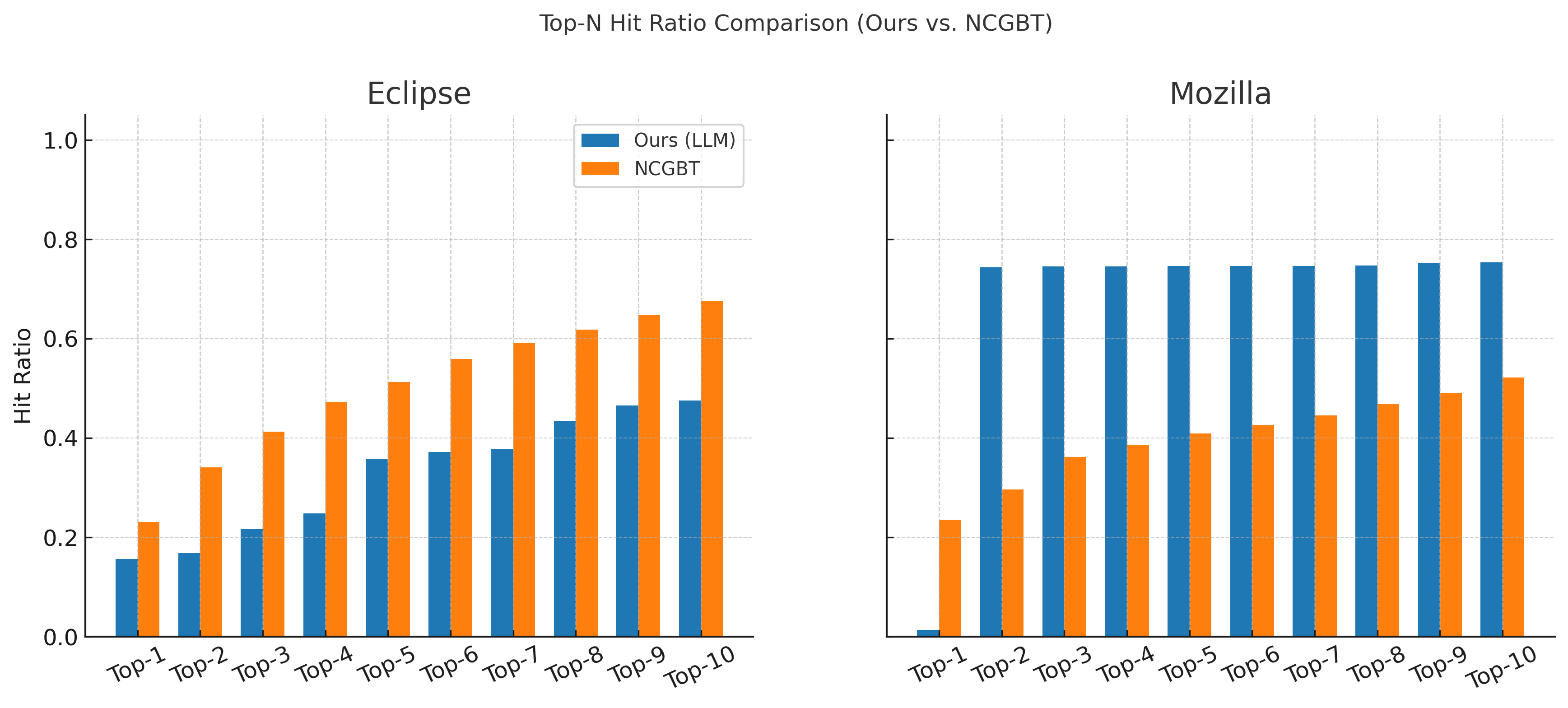}
    \caption{Top-$N$ hit ratio comparison of our LLM-based model and NCGBT~\cite{dong2024neighborhood} on EclipseJDT and Mozilla datasets.}
    \label{fig:topn_bar_comparison}
\end{figure}

Table~\ref{tab:ours_full} reports the full Hit@K statistics for both projects 
EclipseJDT: $n=1{,}612$, Mozilla: $n\approx 11{,}050$
using the multi-year datasets after filtering. 
Figures~\ref{fig:combined_results} and~\ref{fig:topn_bar_comparison} visualize the Hit@K curves.

For an operational view focused on currently active developers, we also report a recent
six-month snapshot; see Sec.~\ref{sec:snapshot}.

\begin{table*}[t]
\centering
\caption{Hit@K for EclipseJDT and Mozilla (multi-year datasets, after filtering). Both counts and ratios are shown.}
\label{tab:ours_full}
\begin{tabular}{lcccccccccc}
\toprule
\multirow{2}{*}{\textbf{Project}} & \multicolumn{10}{c}{\textbf{Hit@K (K=1..10)}}\\
\cmidrule(lr){2-11}
 & 1 & 2 & 3 & 4 & 5 & 6 & 7 & 8 & 9 & 10 \\
\midrule
EclipseJDT (counts) & 251 & 271 & 349 & 400 & 575 & 599 & 610 & 699 & 750 & 765 \\
EclipseJDT (ratio)  & .156 & .168 & .217 & .248 & .357 & .372 & .378 & .434 & .465 & .475 \\
\midrule
Mozilla (counts) & 146 & 8213 & 8227 & 8233 & 8241 & 8243 & 8244 & 8252 & 8308 & 8318 \\
Mozilla (ratio)  & .013 & .743 & .745 & .745 & .746 & .746 & .746 & .747 & .752 & .753 \\
\bottomrule
\end{tabular}
\end{table*}

For EclipseJDT, our model achieves Top-1 = 0.156 and Hit@10 = 0.475. 
For Mozilla, while exact Top-1 accuracy is very low 0.013, the model achieves 
a strong shortlist with Hit@10 = 0.753. These results highlight the challenge of operating 
in the full multi-year label space, where the number of candidate developers is substantially larger 
and long-tail contributors dominate. In particular, the gap between Top-1 and Hit@10 performance can 
be attributed to:  
(i) the extreme sparsity of training examples for many developers (cold-start problem),  
(ii) high variability in bug report style and content across years, and  
(iii) the difficulty of discriminating among multiple plausible assignees under text-only conditioning 
(cf. Sec.~\ref{sec:error-analysis}).

\subsection{Context w.r.t.\ Prior Work}

We contrast our multi-year, text-only results with the NCGBT graph–based method~\cite{dong2024neighborhood}, 
which was evaluated on the same EclipseJDT and Mozilla datasets. While our approach does not exploit 
graph structure, candidate relations, or temporal dynamics, this comparison provides useful context 
for understanding typical Top-1/Top-10 ranges. Our LLM achieves competitive \emph{Hit@10} performance 
on Mozilla but lags behind NCGBT on EclipseJDT, highlighting both the difficulty of operating in a large 
label space and the complementary 
strengths of graph-based versus text-only modeling.

\begin{table}[t]
\centering
\caption{Contextual comparison to NCGBT~\cite{dong2024neighborhood} on multi-year datasets.}
\label{tab:context_ncgbt}
\begin{tabular}{lcccc}
\toprule
\multirow{2}{*}{\textbf{Project}} & \multicolumn{2}{c}{\textbf{Ours (multi-year)}} & \multicolumn{2}{c}{\textbf{NCGBT (multi-year)}} \\
\cmidrule(lr){2-3}\cmidrule(lr){4-5}
 & Top-1 & Top-10 & Top-1 & Top-10 \\
\midrule
EclipseJDT & 0.156 & 0.475 & \textbf{0.2307} & \textbf{0.6752} \\
Mozilla    & 0.013 & \textbf{0.753} & \textbf{0.2356} & 0.5216 \\
\bottomrule
\end{tabular}
\end{table}

As shown in Table~\ref{tab:context_ncgbt}, our model lags NCGBT on EclipseJDT across Top-1 and Top-10, while it surpasses NCGBT on Mozilla at Top-10 (despite very low Top-1).

\begin{table*}[t]
\centering
\caption{Performance comparison of NCGBT and our LLM-based model on EclipseJDT and Mozilla (multi-year datasets).}
\label{tab:ours_vs_ncgbt}
\begin{tabular}{llcccccccccc}
\toprule
\textbf{Datasets} & \textbf{Methods} & \textbf{top-1} & \textbf{top-2} & \textbf{top-3} & \textbf{top-4} & \textbf{top-5} & \textbf{top-6} & \textbf{top-7} & \textbf{top-8} & \textbf{top-9} & \textbf{top-10} \\
\midrule
\multirow{3}{*}{EclipseJDT} 
 & NCGBT & 0.2307 & 0.3406 & 0.4122 & 0.4721 & 0.5121 & 0.5592 & 0.5914 & 0.6175 & 0.6473 & 0.6752 \\
 & \textbf{Ours (LLM)} & 0.156 & 0.168 & 0.217 & 0.248 & 0.357 & 0.372 & 0.378 & 0.434 & 0.465 & 0.475 \\
 & \textbf{Improve} & \textbf{-7.47\%} & \textbf{-17.26\%} & \textbf{-19.52\%} & \textbf{-22.41\%} & \textbf{-15.51\%} & \textbf{-18.72\%} & \textbf{-21.34\%} & \textbf{-18.41\%} & \textbf{-18.08\%} & \textbf{-20.02\%} \\
\midrule
\multirow{3}{*}{Mozilla} 
 & NCGBT & 0.2356 & 0.2964 & 0.3618 & 0.3854 & 0.4085 & 0.4261 & 0.4455 & 0.4681 & 0.4907 & 0.5216 \\
 & \textbf{Ours (LLM)} & 0.013 & 0.743 & 0.745 & 0.745 & 0.746 & 0.746 & 0.746 & 0.747 & 0.752 & 0.753 \\
 & \textbf{Improve} & \textbf{-22.26\%} & \textbf{+44.86\%} & \textbf{+38.32\%} & \textbf{+35.10\%} & \textbf{+33.75\%} & \textbf{+32.49\%} & \textbf{+30.11\%} & \textbf{+27.89\%} & \textbf{+26.15\%} & \textbf{+23.14\%} \\
\bottomrule
\end{tabular}
\end{table*}

\begin{figure}[t]
    \centering
    \includegraphics[width=\linewidth]{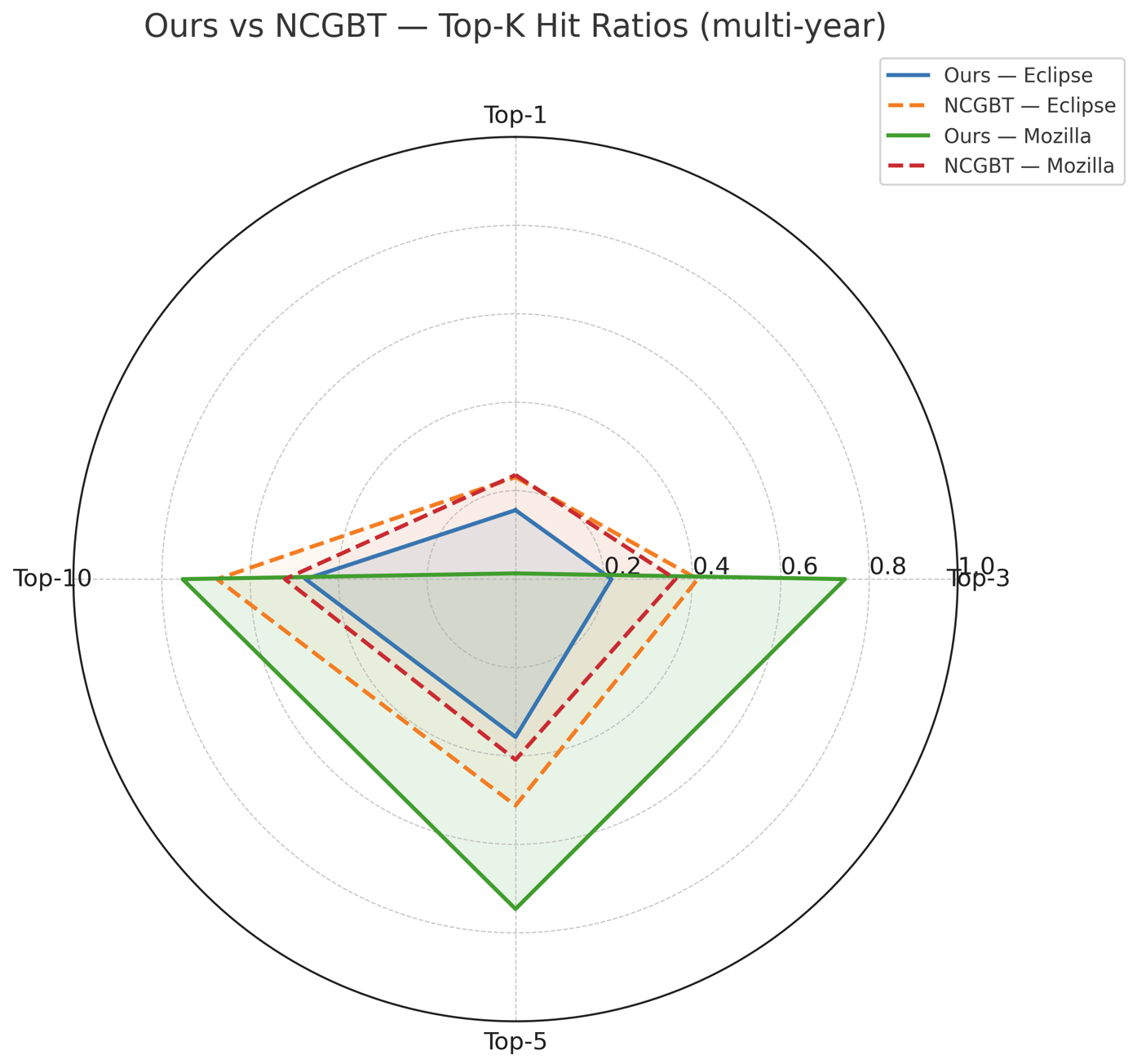}
    \caption{Radar chart comparing}
    \label{fig:radar_comparison}
\end{figure}

With the NCGBT-aligned multi-year evaluation, the picture changes. On EclipseJDT, our LLM trails NCGBT across $K$ (Top-1 $=0.156$ vs.\ $0.2307$, Top-10 $=0.475$ vs.\ $0.6752$). On Mozilla, the model is weak on exact match (Top-1 $=0.013$ vs.\ $0.2356$) but produces substantially stronger shortlists (Top-10 $=0.753$ vs.\ $0.5216$). These outcomes indicate that candidate-constrained decoding helps translate the LLM’s text understanding into \emph{useful ranked recommendations}, yet exact Top-1 assignment remains challenging under a large, long-tail label space. Consistent with prior work, we observe that (i) ranking beyond Top-1 materially improves hit rates, and (ii) project characteristics strongly modulate performance: EclipseJDT appears to favor graph-based modeling of structural/temporal relations, whereas Mozilla benefits more from text-only ranking. The mixed results motivate hybrid approaches that combine LLM ranking with graph-derived priors.

\subsection{Operational Snapshot: Recent Six-Month Window (Jan--Jun 2025)}
\label{sec:snapshot}

While our main evaluation follows the NCGBT multi-year setting, many teams care
most about \emph{recent} activity: who is currently active and who actually fixed
the latest issues. To capture this operational view, we also evaluate on a
six-month snapshot (Jan--Jun 2025) for EclipseJDT and Mozilla using the same
LLM, prompts, training protocol, and candidate-constrained decoding. The label
space is restricted to developers observed in the training split (no test-label
leakage), reflecting how a production system would use a current roster.

\paragraph{Results}
On this recent snapshot, the model attains strong exact-match accuracy on both
projects: EclipseJDT Top-1/Hit@10 = 0.83/0.99 and Mozilla
Top-1/Hit@10 = 0.615/0.72 (each with $n=200$ test issues). These values are not
directly comparable to multi-year metrics due to the much smaller, more stable
label space and reduced drift, but they illustrate the method’s practical utility
when triaging \emph{current} issues.

\begin{table}[t]
\centering
\caption{Six-month snapshot (Jan--Jun 2025): summary results ($n=200$ test per project).}
\label{tab:sixmonth_summary}
\begin{tabular}{lcc}
\toprule
\textbf{Project} & \textbf{Top-1} & \textbf{Hit@10} \\
\midrule
EclipseJDT & 0.830 & 0.990 \\
Mozilla & 0.615 & 0.720 \\
\bottomrule
\end{tabular}
\end{table}

\begin{figure}[t]
  \centering
  \includegraphics[width=\linewidth]{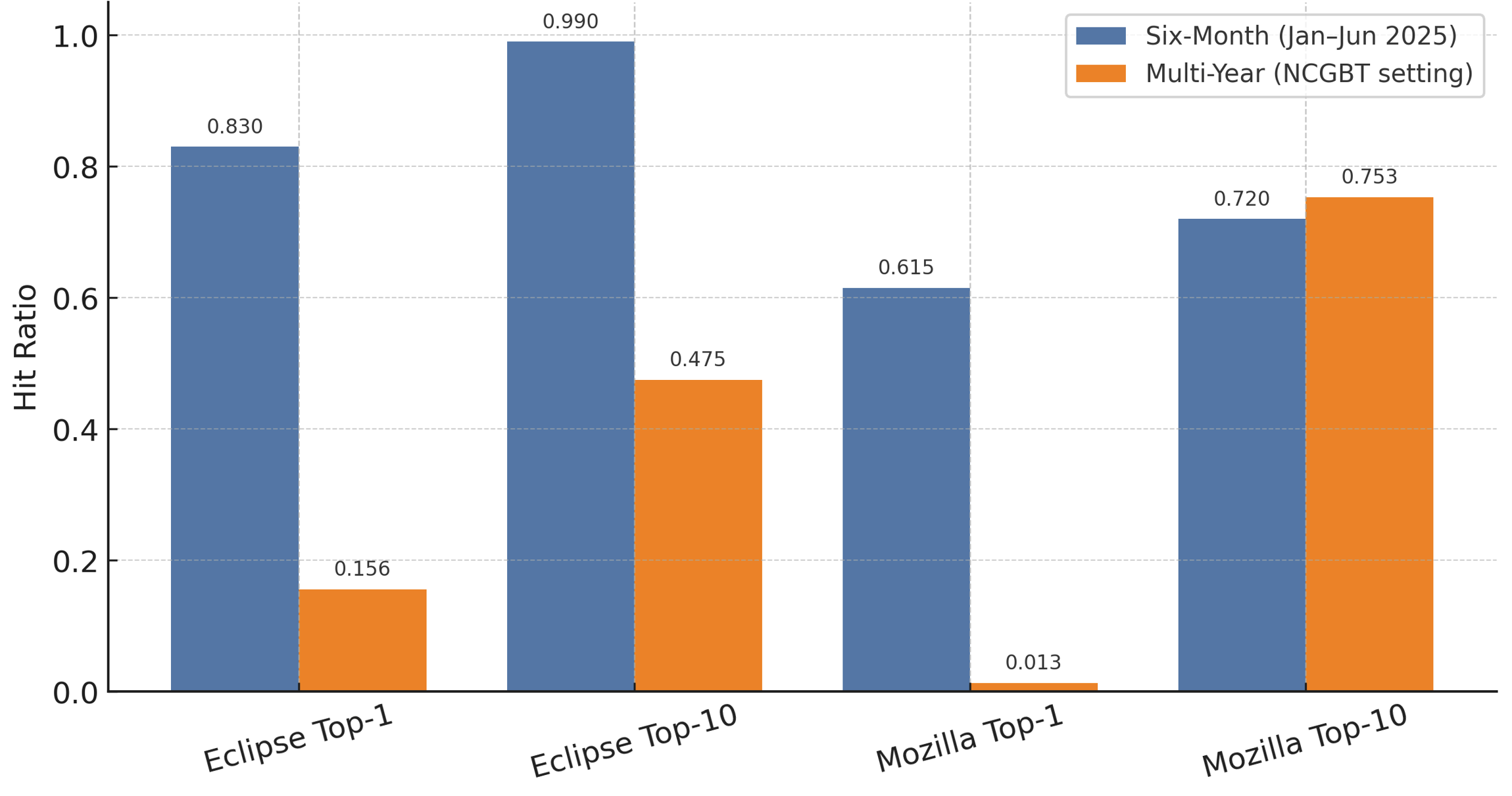}
  \caption{Comparative hit ratios for EclipseJDT and Mozilla under two settings:
  Six-month (Jan--Jun 2025) vs.\ Multi-year (NCGBT setting).}
  \label{fig:window_effect_together}
\end{figure}

\paragraph{Window effect}
Contrasting the six-month snapshot with the multi-year setting highlights how the
label space and turnover shape outcomes. EclipseJDT shows large drops when moving to
the broader, long-tailed label space; Mozilla shows a dramatic Top-1 drop but a
slight \emph{increase} in Hit@10, indicating that shortlist quality can remain high
even as exact matches become harder.

\begin{table}[t]
\centering
\caption{Window effect: six-month (Jan--Jun 2025) vs.\ multi-year (NCGBT setting).}
\label{tab:window_effect}
\begin{tabular}{lcccc}
\toprule
\multirow{2}{*}{\textbf{Project}} & \multicolumn{2}{c}{\textbf{Six-Month}} & \multicolumn{2}{c}{\textbf{Multi-Year}} \\
\cmidrule(lr){2-3}\cmidrule(lr){4-5}
 & Top-1 & Hit@10 & Top-1 & Hit@10 \\
\midrule
EclipseJDT & 0.830 & 0.990 & 0.156 & 0.475 \\
Mozilla    & 0.615 & 0.720 & 0.013 & 0.753 \\
\bottomrule
\end{tabular}
\end{table}

\paragraph{Takeaways}
The six-month view isolates \emph{currently active} developers and reduces the
effective label space, yielding higher Top-1 and near-saturated Top-10 on EclipseJDT
and strong Top-1 on Mozilla. In practice, we recommend reporting both: (i) a
multi-year evaluation for historical completeness and fairness against prior work,
and (ii) a recent snapshot for day-to-day operations (staff changes, departures,
new contributors). This dual reporting makes the method’s deployment value clear
while preserving scientific comparability.

\subsection{Reproducibility}
All experiments were run on a single NVIDIA A100 (80GB) using \texttt{transformers}, \texttt{trl}, \texttt{peft}, and \texttt{bitsandbytes}. We fixed \texttt{seed=3407}, saved intermediate and final LoRA adapters, and published per-issue predictions alongside metrics to enable exact replication.

\subsection{Results and Analysis}

We evaluate on the NCGBT-aligned multi-year datasets (EclipseJDT and Mozilla) using Top-1 accuracy and Hit@$K$ ($K\in\{1,\dots,10\}$). Table~\ref{tab:ours_full} summarizes the full Hit@$K$ statistics for both projects.

Overall, the results show a mixed picture. On EclipseJDT, our LLM underperforms NCGBT across all $K$ (\emph{Top-1} $=0.156$ vs.\ $0.2307$, \emph{Top-10} $=0.475$ vs.\ $0.6752$), indicating that structural/temporal relations captured by graph models remain important for this project. In contrast, on Mozilla the LLM is weak on exact match (\emph{Top-1} $=0.013$ vs.\ $0.2356$) but delivers substantially stronger shortlists (\emph{Top-10} $=0.753$ vs.\ $0.5216$, $\Delta{+}0.231$), suggesting that text-only modeling with candidate-constrained decoding can yield high-quality recommendations even under a very large label space.

Two patterns are consistent across projects: (i) ranking beyond Top-1 materially improves hit rates (monotone Hit@$K$ curves), and (ii) project characteristics modulate outcomes EclipseJDT appears to benefit more from graph structure, whereas Mozilla benefits more from text-driven ranking. These observations motivate hybrid approaches (e.g., graph priors + LLM re-ranking) to improve exact Top-1 without sacrificing shortlist quality.

\subsection{Error Analysis}
\label{sec:error-analysis}

To better understand the limitations of our approach under the multi-year setting, we qualitatively analyzed incorrect or low-ranked predictions.

\paragraph{(E1) Long-tail and cold-start developers.}
A common failure mode involves low-frequency assignees (developers with few resolved bugs). With a large label space and heavy class imbalance, the model often assigns such issues to more frequent but semantically similar developers. This aligns with the observed gap between \emph{Top-1} and \emph{Hit@10} on Mozilla: the true assignee frequently appears in the shortlist but not at rank~1.

\paragraph{(E2) Misranking among plausible candidates.}
On many errors, the generated Top-$K$ list contained multiple reasonable experts. The true developer appeared within the top-$K$ but was \emph{misordered}. This suggests our SFT objective (next-token cross-entropy) optimizes single-label production, not listwise ranking.

\paragraph{(E3) Ambiguous or under-specified reports.}
Some summaries/descriptions lacked concrete signals (component, stack, or repro steps). In these cases the model defaulted to frequent experts, producing reasonable but incorrect assignments.

\paragraph{(E4) Missing structural/temporal cues.}
Compared to NCGBT, our text-only model does not exploit \emph{graph structure} or \emph{temporal recency}. On EclipseJDT where structural/temporal relations are informative this likely contributes to lower Hit@K versus graph-based modeling.

These findings highlight the importance of both data coverage and ranking-aware training. In future work we plan to:
\begin{itemize}
    \item Use few-shot / in-context augmentation for rare developers and class-balanced sampling to mitigate the long tail.
    \item Replace pure SFT with pairwise or listwise ranking losses (e.g., margin or softmax listwise) and/or a lightweight re-ranker trained on Top-$K$ candidates.
    \item Incorporate developer profile embeddings (expertise, components, recency) and temporal priors.
    \item Explore hybrid designs: graph-derived candidate priors + LLM re-ranking with candidate-constrained decoding.
\end{itemize}

Despite these limitations, the model consistently produces \textbf{useful shortlists} (e.g., Mozilla Hit@10 $\approx 0.75$), indicating practical value for human-in-the-loop triaging even when exact Top-1 assignment is challenging.

\section{Threats to Validity}
\label{sec:threats}

We acknowledge several potential threats to the validity of our findings, spanning
internal, external, construct, and conclusion validity.

\subsection{Internal Validity}
All datasets were processed by a uniform pipeline that converts Bugzilla reports
to JSONL with a \texttt{(system, user, assistant)} triplet. Imperfect
tokenization, alias normalization (e.g., email/handle variants), or prompt
templating could bias predictions. To avoid test leakage, candidate rosters were
constructed only from training labels and/or the official assignee lists; we
validated that no test-only identifiers were injected. The model (DeepSeek-R1-Distill-Llama-8B
with LoRA) used fixed seeds and stable hyperparameters; however, single-seed
training can still induce variance, and 4-bit quantization may introduce small
deviations relative to full precision.

\subsection{External Validity}
Evaluation is limited to two large open-source projects (EclipseJDT and Mozilla)
under the \emph{same multi-year temporal windows and developer-filtering protocol}
as NCGBT. Results may not generalize to proprietary trackers, smaller projects,
or ecosystems with different reporting conventions, assignment policies, or label
distributions. Candidate-constrained decoding also assumes access to a reasonably
complete assignee roster; stale or incomplete rosters would degrade performance.
Moreover, the EclipseJDT window ends in 2015 and the Mozilla window in 2021; more
recent projects may exhibit different dynamics.

\subsection{Construct Validity}
We report Top-1 and Hit@$K$ (Top-$N$ hit ratio), which measure whether the true
assignee appears within the top-$K$ predictions. These metrics do not capture
other operational goals (e.g., workload balancing, fairness, or resolution time),
nor do they account for reassignment chains. Noise can arise when the recorded
assignee differs from the actual fixer, or when generic/team accounts are used.
Complementary measures (e.g., MRR/NDCG, calibration, and human-in-the-loop time
savings) are left for future work.

\subsection{Conclusion Validity}
Under the multi-year setting, our LLM underperforms NCGBT on EclipseJDT across
$K$, while producing substantially stronger shortlists on Mozilla (high Hit@10
but weak Top-1). Thus, causal claims about universal superiority are unwarranted.
Observed differences may reflect class imbalance, label-space size, and missing
structural/temporal cues in our text-only model. Statistical testing (e.g.,
bootstrap confidence intervals or Wilcoxon signed-rank on per-issue indicators),
multi-seed runs, and ablations (e.g., with/without candidate constraints or
profile features) would further strengthen the robustness of our conclusions.

\section{Discussion}
\label{sec:disc}
Using the NCGBT-aligned multi-year datasets for EclipseJDT and Mozilla, we evaluated an instruction-tuned LLM (DeepSeek-R1-Distill-Llama-8B) with LoRA fine-tuning and candidate-constrained decoding. The results are mixed across projects: on EclipseJDT, the model attains Top-1 = 0.156 and Hit@10 = 0.475; on Mozilla, it yields a very low Top-1 = 0.013 but a strong Hit@10 = 0.753. This pattern suggests that, under large label spaces and heavy long tails, the LLM produces \emph{useful ranked shortlists} even when exact Top-1 assignment is challenging.

In direct comparison with the graph-based NCGBT~\cite{dong2024neighborhood} on the same datasets, our LLM \emph{lags} on EclipseJDT across $K$ but delivers substantially stronger shortlists on Mozilla (higher Hit@10). We hypothesize that EclipseJDT benefits more from structural/temporal relations captured by graph modeling, whereas Mozilla’s heterogeneous report text and broader label space allow text-driven ranking to surface the correct assignee within the top-$K$ more frequently.

A central factor behind shortlist quality is candidate-constrained decoding, which prevents hallucinated identifiers and aligns predictions with the valid roster. Consistent prompt formatting also stabilizes instruction following. From an efficiency standpoint, 4-bit NF4 quantization with LoRA adapters enables practical fine-tuning and inference on limited hardware, supporting continuous deployment.

\paragraph{Operational implications.}
Even when Top-1 is low, a high Hit@10 can materially reduce triage effort in \emph{human-in-the-loop} workflows (e.g., presenting a ranked shortlist to maintainers). Integrating our model as a recommendation panel in the issue tracker or CI/CD bots can accelerate assignment while preserving human oversight.

\paragraph{Practical utility despite low Top-1.}
Although our model often underperforms on exact Top-1 accuracy in the multi-year setting, the high Hit@$K$ values (e.g., 0.753 on Mozilla at $K=10$) demonstrate tangible utility in practice. In real bug triaging workflows, maintainers rarely expect perfect automated assignment; instead, they benefit from a shortlist of plausible candidates that reduces search space and decision effort. For example, presenting the top 5--10 developers can allow a triager to make an informed assignment within seconds rather than manually scanning project history or guessing based on incomplete context. This is especially valuable under large, long-tailed label spaces where exact assignment is inherently difficult. Thus, even with modest Top-1 accuracy, our method can significantly lower human triaging time, increase consistency, and serve as a reliable assistive tool rather than a fully autonomous replacement.

\paragraph{Where to go next.}
The mixed outcomes motivate \emph{hybrid} designs: (i) use graph-derived priors or developer profile embeddings (expertise, components, recency) to inform candidate sets, and (ii) apply pairwise/listwise objectives or a lightweight re-ranker to improve \emph{intra-list} ordering. Retrieval-augmented candidate expansion and cross-project transfer are promising for cold-start developers. Finally, multi-seed runs, ablations (with/without constraints or profiles), and calibration analysis would further strengthen conclusions and improve exact-match performance.

\subsection{Future Directions and Developer Embeddings}

In this study, the model relied solely on the textual content of Bugzilla issues
(\emph{title + body}) with candidate-constrained decoding at inference. While this
text-only setup produced \emph{useful shortlists} under large label spaces (e.g.,
Mozilla Hit@10 $\approx 0.75$), exact Top-1 assignment remained challenging and
performance on EclipseJDT lagged graph-based baselines.

A promising extension is the integration of \textit{developer profile embeddings}
that encode historical activity, component ownership, and prior bug-fix patterns.
Concretely, we envisage:
\begin{itemize}
  \item Profile features: per-developer embeddings from resolved issues
        (components, keywords, files, temporal recency), with alias normalization.
  \item Retrieval + re-ranking: a two-stage pipeline where issues retrieve
        top candidates via similarity to developer profiles, followed by LLM
        re-ranking under candidate constraints.
  \item Listwise objectives: augment SFT with pairwise/listwise losses
        to improve intra-list ordering (Top-$K$ ranking quality).
  \item Temporal priors: decay factors for recent activity to mitigate
        stale expertise and improve Top-1.
  \item Hybrid cues: optionally inject graph-derived signals (e.g.,
        component co-activity) as soft priors or prompt descriptors.
\end{itemize}
We expect these additions to enhance Top-$K$ accuracy, reduce sensitivity to class
imbalance, and better address cold-start scenarios.

\section{Ethical Considerations}

This study utilizes real bug reports and developer identifiers (e.g., names and email addresses) sourced from publicly available issue trackers. Although the data is publicly accessible, we recognize the importance of safeguarding contributor privacy and reducing the risk of unintended exposure. In any future deployment or public release, all developer identifiers will be anonymized to prevent potential misuse or profiling.

Our model, trained on historical issue assignments, may reflect and potentially amplify existing biases in the underlying data. For example, it may disproportionately favor highly active or senior developers, thereby reinforcing workload imbalances. This risk is particularly pronounced in datasets with skewed assignment histories. To address such concerns, we recommend integrating fairness-aware training objectives, distribution-aware sampling strategies, or post-hoc bias mitigation techniques to promote equitable assignment in production environments.

From a sustainability standpoint, large language models generally require substantial computational resources. To reduce environmental impact and operational cost, we employed parameter-efficient fine-tuning via LoRA combined with 4-bit NF4 quantization, significantly lowering GPU memory usage and power consumption during both training and inference. Nevertheless, we acknowledge that LLM-based systems still contribute to carbon emissions. Future work should explore energy-efficient architectures, carbon accounting mechanisms, and renewable-powered training infrastructure to further advance responsible AI development.
Our approach aligns with broader efforts in software security where LLMs are increasingly leveraged for vulnerability detection and secure coding practices, demonstrating their growing role in practical software reliability~\cite{kiashemshaki2025secure}.

\section{Conclusion and Future Work}
\label{sec:concl}

We introduced an LLM-based framework for automated issue triaging that fine-tunes
\emph{DeepSeek-R1-Distill-Llama-8B} with parameter-efficient LoRA adapters on real-world
Bugzilla data, using a conversational JSONL format and 4-bit NF4 quantization for
compute efficiency. At inference, \emph{candidate-constrained decoding} restricts outputs
to a valid assignee roster, preventing hallucinated identifiers and producing ranked
Top-$K$ recommendations.

Evaluated on the \emph{same multi-year temporal windows and filtering protocol} as NCGBT,
the results are mixed across projects: on EclipseJDT, our text-only model lags behind the
graph-based baseline across $K$; on Mozilla, it delivers \emph{strong shortlists} (high
Hit@10) despite low Top-1 accuracy. These outcomes indicate that instruction-tuned LLMs
can provide practical, human-in-the-loop recommendations under large label spaces, while
exact Top-1 assignment remains challenging without structural/temporal signals.

Operationally, the framework is lightweight minimal preprocessing, no handcrafted
graph construction, and modest hardware demands making it suitable for integration into
issue trackers or CI/CD bots. Periodic roster refresh and incremental fine-tuning enable
adaptation to evolving teams and components.

Future work will target accuracy, robustness, and deployment utility:
\begin{itemize}
    \item Developer embeddings \& temporal priors: encode expertise, component ownership, and recency to improve disambiguation and Top-1.
    \item Retrieval + re-ranking: two-stage pipelines that retrieve candidate developers via profile similarity, then apply LLM re-ranking under candidate constraints.
    \item Ranking-aware training: pairwise/listwise losses or a lightweight re-ranker to improve intra-list ordering (MRR/NDCG, not just Hit@$K$).
    \item Hybrid models: inject graph-derived signals as priors while retaining the LLM’s text understanding.
    \item Human-in-the-loop tooling: confidence scores, abstention, triager feedback loops, and workload-aware re-ranking for fair assignment.
\end{itemize}

Similar to recent work where LLMs serve as adaptive control agents in complex systems like wireless body area networks~\cite{torkamani2025llm}, future bug triaging systems could evolve toward dynamic, context-aware assignment strategies.

In sum, instruction-style fine-tuning plus candidate-constrained decoding is a practical,
scalable foundation for automated triage: it already yields useful Top-$K$ shortlists in
large, long-tailed settings, and offers a clear path via profile/graph augmentation and
ranking-aware objectives to stronger exact-match performance.

\bibliographystyle{IEEEtran}
\bibliography{mybib}

\end{document}